\documentclass[aps,pra,twocolumn,amsmath,amssymb,floatfix]{revtex4-1}

\usepackage[USenglish]{babel}
\usepackage[utf8]{inputenc}
\usepackage{graphicx}
\usepackage{xparse}
\usepackage{etoolbox}
\usepackage{mathtools}
\usepackage{subdepth}
\usepackage{physics}
\usepackage{hyperref}
\usepackage{glossaries}
\usepackage[capitalise]{cleveref}
\usepackage{csquotes}
\usepackage[abbreviations,USenglish]{foreign}
\usepackage{dsfont}
\usepackage{xparse}
\usepackage{tabularx}

\makeatletter
  \usepackage{xspace}
  \usepackage{xpunctuate}

  \renewcommand\UKUS@comma{\xperiodcommaafter}

  \appto{\appendix}{%
    \@ifstar{\def\theequation@prefix{A.}}%
            {}}
\makeatother

\newcommand{\imag}{\mathrm{i}\mkern1mu} % see ISO 80000-2:2009
\let\vec\vectorbold%

\renewcommand\toprule{\hline\hline}
\newcommand\midrule{\hline}
\newcommand\bottomrule{\hline\hline}

\newcolumntype{C}{>{\centering\arraybackslash}X}

\crefname{section}{Sec.}{Section}
\crefname{pluralequation}{Eqs.}{Equations}
\Crefname{pluralequation}{Eqs.}{Equations}

\creflabelformat{equation}{#2(#1)#3}
\creflabelformat{pluralequation}{#2(#1)#3}

\crefalias{subequation}{equation}
\crefalias{subsection}{section}

\NewDocumentCommand\subrefformat{m}{(#1)}
\NewDocumentCommand\subrefsformat{mmg}{%
  \IfNoValueTF{#3}{(#1) and (#2)}{(#1), (#2), and (#3)}}
\NewDocumentCommand\subrefrangeformat{mm}{(#1)--\allowbreak(#2)}

\NewDocumentCommand\csubref{smm}{%
  \IfBooleanTF{#1}{\cref*{#2}}{\cref{#2}}~\subrefformat{#3}%
}
\NewDocumentCommand\Csubref{smm}{%
  \IfBooleanTF{#1}{\Cref*{#2}}{\Cref{#2}}~\subrefformat{#3}%
}
\NewDocumentCommand\labelcsubref{smm}{%
  \IfBooleanTF{#1}{\labelcref*{#2}}{\labelcref{#2}}~\subrefformat{#3}%
}

\NewDocumentCommand\csubrefs{smmgg}{%
  \namecrefs{#2}%
  ~\IfBooleanTF{#1}{\labelcref*{#2}}{\labelcref{#2}}%
  ~\IfNoValueTF{#4}
  {\subrefformat{#3}}
  {\IfNoValueTF{#5}
    {\subrefsformat{#3}{#4}}
    {\subrefsformat{#3}{#4}{#5}}}%
}
\NewDocumentCommand\Csubrefs{smmgg}{%
  \nameCrefs{#2}%
  ~\IfBooleanTF{#1}{\labelcref*{#2}}{\labelcref{#2}}%
  ~\IfNoValueTF{#4}
  {\subrefformat{#3}}
  {\IfNoValueTF{#5}
    {\subrefsformat{#3}{#4}}
    {\subrefsformat{#3}{#4}{#5}}}%
}
\NewDocumentCommand\labelcsubrefs{smmgg}{%
  \IfBooleanTF{#1}{\labelcref*{#2}}{\labelcref{#2}}%
  ~\IfNoValueTF{#4}
  {\subrefformat{#3}}
  {\IfNoValueTF{#5}
    {\subrefsformat{#3}{#4}}
    {\subrefsformat{#3}{#4}{#5}}}%
}

\NewDocumentCommand\csubrefrange{smmm}{%
  \namecrefs{#2}%
  ~\IfBooleanTF{#1}{\labelcref*{#2}}{\labelcref{#2}}%
  ~\subrefrangeformat{#3}{#4}%
}
\NewDocumentCommand\Csubrefrange{smmm}{%
  \nameCrefs{#2}%
  ~\IfBooleanTF{#1}{\labelcref*{#2}}{\labelcref{#2}}%
  ~\subrefrangeformat{#3}{#4}%
}
\NewDocumentCommand\labelcsubrefrange{smmm}{%
  \IfBooleanTF{#1}{\labelcref*{#2}}{\labelcref{#2}}%
  ~\subrefrangeformat{#3}{#4}%
}

\usepackage{tensor}

\NewDocumentCommand\bileft{ m }{\mathopen{}\mathclose\bgroup\left#1}
\NewDocumentCommand\biright{ m }{\aftergroup\egroup\right#1}
\NewDocumentCommand\bitensor{ o m o }
{\left.\kern-\nulldelimiterspace\IfNoValueTF{#1}
  {
    \IfNoValueTF{#3}
    {\tensor*[]{#2}{}}
    {\tensor*[]{#2}{_{\mathrm{#3}}}}
  }{
    \IfNoValueTF{#3}
    {\tensor*[_{\mathrm{#1}}]{#2}{}}
    {\tensor*[_{\mathrm{#1}}]{#2}{_{\mathrm{#3}}}}
  }\right.\kern-\nulldelimiterspace}

\NewDocumentCommand\biket{ m s m }
{ % Ket
  \bitensor{
    \IfBooleanTF{#2}
    {\vphantom{#3}\bileft\lvert\smash{#3}\biright\rangle} % No resize
    {\bileft\lvert{#3}\biright\rangle} % Auto sizing
  }[#1]
}

\NewDocumentCommand\lket{}{\biket{L}}
\NewDocumentCommand\rket{}{\biket{R}}

\NewDocumentCommand\bibra{ m s m }
{ % Bra
  \bitensor[#1]{
    \IfBooleanTF{#2}
    {\vphantom{#3}\bileft\langle\smash{#3}\biright\rvert} % No resize
    {\bileft\langle{#3}\biright\rvert} % Auto sizing
  }
}

\NewDocumentCommand\lbra{}{\bibra{L}}
\NewDocumentCommand\rbra{}{\bibra{R}}

\NewDocumentCommand\bibraket{ m m s m g g }
{ % Inner product
  \bitensor[#1]{
    \IfBooleanTF{#3}
    { % No resize
      \IfNoValueTF{#5}
      {\vphantom{#4}\langle\smash{#4}\vert\smash{#4}\biright\rangle}
      {
        \IfNoValueTF{#6}
        {\vphantom{#4#5}\langle\smash{#4}\vert\smash{#5}\rangle}
        {\vphantom{#4#5#6}\langle\smash{#4}\vert\smash{#5}\vert\smash{#6}\rangle}
      }
    }
    { % Auto resize based on bra/ket arguments
      \IfNoValueTF{#5}
      {\bileft\langle{#4}\middle\vert{#4}\biright\rangle}
      {
        \IfNoValueTF{#6}
        {\bileft\langle{#4}\middle\vert{#5}\biright\rangle}
        {\vphantom{#5}\bileft\langle{#4}\middle\vert\smash{#5}\middle\vert{#6}\biright\rangle}
      }
    }
  }[#2]
}

\NewDocumentCommand\lrbraket{}{\bibraket{L}{R}}

\NewDocumentCommand\biketbra{ m m s m g g }
{ % Dyadic product
  \bitensor[#1]{
    \IfBooleanTF{#3}
    { % No resize
      \IfNoValueTF{#5}
      {\vphantom{#4}\lvert\smash{#4}\rangle\!\langle\smash{#4}\rvert}
      {
        \IfNoValueTF{#6}
        {\vphantom{#4}\lvert\smash{#4}\rangle\!\langle\smash{#5}\rvert}
        {\vphantom{#4#5#6}\lvert\smash{#4}\rangle\smash{#5}\langle\smash{#6}\rvert}
      }
    }
    { % Auto resize based on bra/ket arguments
      \IfNoValueTF{#5}
      {\bileft\lvert{#4}\middle\rangle\!\middle\langle{#4}\biright\rvert}
      {
        \IfNoValueTF{#6}
        {\bileft\lvert{#4}\middle\rangle\!\middle\langle{#5}\biright\rvert}
        {\vphantom{#5}\bileft\lvert{#4}\middle\rangle\smash{#5}\middle\langle{#6}\biright\rvert}
      }
    }
  }[#2]
}

\NewDocumentCommand\llketbra{}{\biketbra{L}{L}}

\NewDocumentCommand\rrketbra{}{\biketbra{R}{R}}

\NewDocumentCommand\lketbra{}{\llketbra}
\NewDocumentCommand\rketbra{}{\rrketbra}

\begin{document}

% Use the \preprint command to place your local institutional report
% number in the upper righthand corner of the title page in preprint mode.
% Multiple \preprint commands are allowed.
% Use the 'preprintnumbers' class option to override journal defaults
% to display numbers if necessary
% \preprint{}

\begin{abstract}
The experimental realization of balanced gain and loss in a quantum system has been a long standing goal in quantum mechanics since the introduction of the concept of $\mathcal{PT}$ symmetry and has only recently been achieved. In this paper we analyze balanced gain and loss in Gaussian multi-well potentials with either only gain or loss in each well. By means of symmetrization via matrix models we can construct asymmetric extended potentials with partially real or complex conjugate spectra. This will be demonstrated explicitly for double-well and triple-well systems. Such systems can be realized with Bose-Einstein condensates in optical trapping potentials in the presence of localized particle gain and loss. The usage of asymmetric potentials in the process is more versatile and is considered beneficial in real experimental implementations.
\end{abstract}

\def\itpeins{\affiliation{Institut f\"ur Theoretische Physik 1, Universit\"at Stuttgart, 70550 Stuttgart, Germany}}

\title{Balanced gain and loss in spatially extended non-\texorpdfstring{$\mathcal{PT}$}{PT}-symmetric multi-well potentials}
\author{Sinan Altinisik}
\email{sinan.altinisik@itp1.uni-stuttgart.de}
\itpeins

\author{Daniel Dizdarevic}
\author{Jörg Main}\itpeins

\date{\today}
\keywords{keywords}
\maketitle

\section{Introduction}
\label{subsec:introduction}

It is well-known that gain and loss in open quantum systems can effectively be described by complex potentials \cite{Graefe2008}. If gain and loss are balanced, then the corresponding non-Hermitian Hamiltonian possesses real energy eigenvalues. Great interest in non-Hermitian quantum mechanics arose with the introduction of the concept of $\mathcal{PT}$ symmetry by Bender and Boettcher \cite{Bender1998}. The spectrum of a $\mathcal{PT}$-symmetric quantum system consists of real and pairs of complex conjugate energy eigenvalues. This means that gain and loss can be balanced in a $\mathcal{PT}$-symmetric quantum system.

However, the occurrence of real and pairs of complex conjugate energy eigenvalues in the spectrum of a non-Hermitian Hamiltonian was already discovered earlier within the more general framework of quasi-Hermiticity \cite{Scholtz1992}. While a potential of a $\mathcal{PT}$-symmetric quantum system must possess a symmetric real and an antisymmetric imaginary part, a quasi-Hermitian quantum system on the other hand allows for completely asymmetric potentials.

Nevertheless, it were $\mathcal{PT}$-symmetric systems which were first proposed \cite{Ruschhaupt2018} and realized \cite{Guo2009} experimentally in optical systems a decade after the introduction of the concept. Since then, numerous other experiments and applications of balanced gain and loss in classical $\mathcal{PT}$-symmetric and quasi-Hermitian systems have been reported \cite{Schindler2011,Schindler2012,Ramezani2012,Bender2013,Chong2011,Ge2011,Liertzer2012,Sternheim1972,Ruschhaupt2018,Kreibich2016,Makris2015,Brandsttter2019,Rivet2018}. However, it took another decade until the first observations of $\mathcal{PT}$ symmetry in different quantum systems were made \cite{Li2019,Wu2019,Naghiloo2019}. Yet another promising candidate for an experimental realization is a Bose-Einstein condensate in a multi-well optical potential with localized particle gain and loss as proposed in Ref.~\cite{Klaiman2008}. For Bose-Einstein condensates one can create arbitrary optical potentials in time average \cite{Henderson2009} and also the experimental realization of localized loss \cite{Gericke2008,Wrtz2009,Barontini2013} and of localized gain \cite{Dring2009} is possible. The advantages of this approach are twofold: First, the mathematics involved to describe a Bose-Einstein condensate in a multi-well potential is the same as for the description of a large class of systems, among which are optical systems with a Kerr nonlinearity \cite{Agrawal2001,Morsch2006,Ramezani2010}, polarons \cite{Holstein1959a,Holstein1959b,Campbell1982}, and excitons \cite{Toyozawa1983}. Second, such systems offer a large amount of control, i.\,e.\ one can investigate almost arbitrary complex potentials.

Most works on the topic of balanced gain and loss in Bose-Einstein condensates use $\mathcal{PT}$-symmetric systems \cite{Robins2008,Kreibich2013,Kreibich2014,Kreibich2016,Kogel2019,Guthrlein2015}. $\mathcal{PT}$ symmetry, although simple from a theoretical point of view, has the restriction that the parameters of the complex potential have to be simultaneously adjusted very precisely. This is demanding, in particular due to the challenging realization of localized gain \cite{Dring2009}. An interesting approach to avoid this problem was presented in Ref.~\cite{Lunt2017}, where a small asymmetry of the potential was stabilized by the nonlinearity of the Gross-Pitaevskii equation arising from the contact interaction between the atoms.

In this paper, however, we want to use a systematic approach and exploit the concept of symmetrization \cite{Dizdarevic2019}, which allows for the construction of asymmetric potentials with balanced gain and loss. Symmetrization has already been used successfully within the framework of matrix models \cite{Dizdarevic2019}, where a whole range of potentials with balanced gain and loss was found. In these matrix models it is possible to take almost arbitrary values for some of the parameters and obtain balanced gain and loss by adjusting the remaining ones. This comes in handy if, for example, there are some potential parameters which are hard to control in an experiment. This is clearly not possible for $\mathcal{PT}$-symmetric systems, where all the potential parameters have to be chosen exactly to fulfill the required symmetries.

The goal of this paper now is to transfer previous results from the matrix model in Ref.~\cite{Dizdarevic2019} to a continuous system, that is, an asymmetric complex multi-well potential in the form of Gaussian functions with either only gain or loss in each well. Our goal is to determine the potential parameters in such a way that the eigenvalues become real or emerge in complex conjugate pairs. If the potential wells are strongly localized, then the system can be well approximated by a matrix model. Therefore, we search for spatially extended potentials which correspond to symmetrized matrix models. We expect that such systems then possess at least the same number of real and complex conjugate energy eigenvalues as the matrix model.

The paper is organized as follows. In \cref{subsec:bgl} we will introduce Bose-Einstein condensates with balanced gain and loss which will be described by a non-Hermitian Hamiltonian. Then we will take a short look at the circumstances under which a non-Hermitian Hamiltonian can posses real eigenvalues in \cref{subsec:non_herm}. \Cref{subsec:method} will deal with the construction of multi-well potentials yielding real or pairs of complex conjugate eigenvalues. This is achieved by means of varying the parameters of the extended potential in such a way that it corresponds to a symmetrized matrix model. Afterwards, the potential parameters are varied again numerically until the eigenvalues are real. In \cref{sec:results} we will present the results for double-well and triple-well potentials. \Cref{sec:conclusions} will finally summarize the contents of this paper and give a short outlook on open questions.

\section{Theory}%
\label{sec:theory}

\subsection{Balanced gain and loss for Bose-Einstein condensates}%
\label{subsec:bgl}

In the mean-field limit Bose-Einstein condensates can be well described by the non-Hermitian Gross-Pitaevskii equation \cite{Dalfovo1999} in dimensionless units ($\hbar = m = 1$),
\begin{equation}
  \imag \pdv{t} \psi(\vec{r},t)
  = \qty(-\nabla^2 + V\qty(\vec{r})
      + g \abs{\psi\qty(\vec{r}, t)}^2) \psi(\vec{r},t) ,
  \label{eq:time_dep_SE}
\end{equation}
which corresponds the mean-field approximation of a quantum master equation \cite{Dast2014}. Most of the atoms in the condensate are then condensed into the same state described by the macroscopic wave function $\psi(\vec{r},t)$. The particle density of the condensate is given by $n(\vec{r},t) = \abs{\psi(\vec{r},t)}^2$. The nonlinearity $g$ arising from the contact interaction between the particles can be tuned via Fesh\-bach resonances \cite{Inouye1998,Pollack2009} within a large range of values including the linear case $g=0$. As from a mathematical point of view the linear case is far easier to treat than the nonlinear case, we will limit our further analysis to non-interacting condensates with $g = 0$.

An effective description of gain and loss in a Bose-Einstein condensate can be achieved with complex potentials, in which the imaginary part $V_\imag(\vec{r})$ plays the role of source and drain of the particle density. Thus, the overall particle number $\mathcal{N}(t)$ is not conserved and changes as
\begin{equation}
  \dot{\mathcal{N}}(t)
  = \int \dd[3]{r} 2 V_\mathrm{i}\qty(\vec{r}) n\qty(\vec{r}, t) .
  \label{eq:continuity_eq_int}
\end{equation}
The particle number increases with positive and decreases with negative expectation values of the imaginary part of the potential. Gain and loss are balanced if $\dot{\mathcal{N}} = 0$.

If we consider stationary solutions of \cref{eq:time_dep_SE} in the form $\psi\qty(\vec{r}, t)=\mathrm{e}^{-\mathrm{i}\mu t}\phi\qty(\vec{r})$ with the chemical potential $\mu\in\mathbb{C}$ and with $n\qty(\vec{r}, t)=\mathrm{e}^{2\Im\mu t}\abs{\phi\qty(\vec{r})}^2$, we find
\begin{equation}
    \Im\mu=\int \dd[3]{r} V_\text{i}(\vec{r})|\phi(\vec{r})|^2.
    \label{eq:imag_part_of_mu}
\end{equation}
This shows that gain and loss are balanced if $\mu\in\mathbb{R}$ which gives rise to the question whether the complex potential $V\qty(\vec{r})$ allows for real or pairs of complex conjugate eigenvalues. However, in order to answer this question we will shortly summarize under which circumstances a general non-Hermitian Hamiltonian can have real eigenvalues.

\subsection{Symmetrization in non-Hermitian quantum mechanics}%
\label{subsec:non_herm}

We consider right and left eigenstates of a non-Hermitian Hamiltonian $\hat{H}$ defined by
\begin{subequations}%
  \label[pluralequation]{subeqs:right_left}
  \begin{align}
    \hat{H} \rket{\psi_n} &= \mu_n \rket{\psi_n} , \label{eq:right_ket} \\
    \lbra{\psi_n} \hat{H} &= \lbra{\psi_n} \mu_n . \label{eq:left_bra}
  \end{align}
\end{subequations}%
While for a Hermitian Hamiltonian right and left eigenstates are equal, this is in general not the case for non-Hermitian Hamiltonians. However, the eigenstates of two non-degenerate discrete eigenvalues $E_n$ and $E_m$ are orthogonal in the sense that
\begin{equation}
  \lrbraket{\psi_m}{\psi_n} = \delta_{mn} .
  \label{eq:ortho}
\end{equation}
If they additionally fulfill the completeness relation
\begin{equation}
  \sum_n \rket{\psi_n} \lbra{\psi_n} = \mathds{1},
  \label{eq:compl}
\end{equation}
then they form a complete bi-orthonormal basis \cite{Mostafazadeh2002}. We want to emphasize, however, that not every non-Hermitian Hamiltonian admits such a basis.
%However, we limit our analysis to Hamiltonians with non-degenerate discrete spectra whose eigenstates form a complete bi-orthonormal basis.

In general, a non-Hermitian Hamiltonian has complex eigenvalues. A criterion for the occurrence of real eigenvalues is symmetrizability \cite{Dizdarevic2019}. A Hamiltonian $\hat{H}$ is called symmetrizable if there exists a pair of linear Hermitian operators $\hat{\eta}_\mathrm{L}$ and $\hat{\eta}_\mathrm{R}$ so that
\begin{subequations}%
  \label[pluralequation]{subeqs:quasi_herm}
  \begin{align}
    \hat{\eta}_\mathrm{L} \hat{H} &= \hat{H}^\dagger \hat{\eta}_\mathrm{L} ,
    \label{eq:quasi_herm_L} \\
    \hat{\eta}_\mathrm{R} \hat{H}^\dagger &= \hat{H} \hat{\eta}_\mathrm{R} .
    \label{eq:quasi_herm_R}
  \end{align}
\end{subequations}
We can then show that
\begin{subequations}%
  \label[pluralequation]{subeqs:yang_to_real}%
  \begin{align}
    \rbra{\psi_n} \hat{\eta}_\mathrm{L}^\dagger \hat{H}
      &= \rbra{\psi_n} \hat{H}^\dagger \hat{\eta}_\mathrm{L}^\dagger
      = \rbra{\psi_n} \hat{\eta}_\mathrm{L}^\dagger \mu_n^\ast ,
      \label{eq:yang_to_real1} \\
    \hat{H} \hat{\eta}_\mathrm{R} \lket{\psi_n}
      &= \hat{\eta}_\mathrm{R} \hat{H}^\dagger \lket{\psi_n}
      = \mu_n^\ast \hat{\eta}_\mathrm{R} \lket{\psi_n} ,
      \label{eq:yang_to_real2}
  \end{align}
\end{subequations}%
which means that if $\mu_n$ is an eigenvalue of $\hat{H}$, so is $\mu_n^\ast$, as long as the corresponding eigenstates $\rket{\psi_n}$ and $\lket{\psi_n}$ are not in the kernels of the symmetrization operators $\hat{\eta}_\mathrm{R}$ and $\hat{\eta}_\mathrm{L}$, respectively. In the literature the conditions \labelcref{subeqs:quasi_herm} are presented with many different names \cite{Scholtz1992,Dizdarevic2019,Mostafazadeh2002,Znojil2008,Nixon2016}, differing mainly in the properties of the operators $\hat{\eta}_\mathrm{L}$ and $\hat{\eta}_\mathrm{R}$. At this point we will not make any further assumptions about such properties.

Assuming that the spectrum is non-degenerate, \cref{subeqs:right_left} yield the following relations between right and left side eigenstates,
\begin{subequations}%
  \label[pluralequation]{subeqs:connection}%
  \begin{align}
    \hat{\eta}_\mathrm{L} \rket*{\psi_{n_0}} &= \lket*{\psi_{n_0}} ,
      \label{eq:connection1} \\
    \hat{\eta}_\mathrm{L} \rket*{\psi_{n_\pm}} &= \lket*{\psi_{n_\mp}} ,
      \label{eq:connection2} \\
    \hat{\eta}_\mathrm{R} \lket*{\psi_{n_0}} &= \rket*{\psi_{n_0}} ,
      \label{eq:connection3} \\
    \hat{\eta}_\mathrm{R} \lket*{\psi_{n_\pm}} &= \rket*{\psi_{n_\mp}} ,
      \label{eq:connection4}
    \end{align}
\end{subequations}
where the indexes $n_0$, $n_+$, and $n_-$ denote eigenstates with real and pairs of complex conjugate energies, respectively.

With \cref{subeqs:connection} we can derive representations of the operators $\hat{\eta}_\mathrm{L}$ and $\hat{\eta}_\mathrm{R}$ in terms of the eigenstates of $\hat{H}$,
\begin{subequations}%
  \label[pluralequation]{subeqs:etas}
  \begin{align}
    \hat{\eta}_\mathrm{L}
      &= \sum_{n_0} \lketbra*{\psi_{n_0}}{\psi_{n_0}} \nonumber \\
      &\quad + \sum_{n_+} \qty(\lketbra*{\psi_{n_-}}{\psi_{n_+}}
        + \lketbra*{\psi_{n_+}}{\psi_{n_-}}) ,
      \label{eq:etal}\\
    \hat{\eta}_\mathrm{R}
      &= \sum_{n_0} \rketbra*{\psi_{n_0}}{\psi_{n_0}} \nonumber \\
      &\quad + \sum_{n_+} \qty(\rketbra*{\psi_{n_-}}{\psi_{n_+}}
        + \rketbra*{\psi_{n_+}}{\psi_{n_-}}) ,
      \label{eq:etar}
  \end{align}
\end{subequations}
where the sums run over all states not being in the kernels of $\hat{\eta}_\mathrm{L}$ or $\hat{\eta}_\mathrm{R}$. If the kernels of $\hat{\eta}_\mathrm{L}$ and $\hat{\eta}_\mathrm{R}$ are empty we say that $\hat{H}$ is symmetrizable, otherwise $\hat{H}$ is only semi-symmetrizable. All eigenstates of the Hamiltonian which are not in the kernels of the symmetrization operators \labelcref{subeqs:etas} correspond to real or pairs of complex conjugate eigenvalues.

We now want to describe Bose-Einstein condensates by a Schr\"odinger equation in position space. The corresponding Hamiltonian has the form $\hat{H} = \vec{p}^2 + V(\vec{x})$ and thus satisfies $\hat{H}^\dagger = \hat{H}^\ast = \vec{p}^2 + V^\ast(\vec{x})$. The right and left eigenvalue equations can thus be written as
\begin{subequations}%
  \label[pluralequation]{subeqs:SSE_PS}
  \begin{align}
    \hat{H} \psi_{n, \mathrm{R}}(x)
      &= \mu_n \psi(x)_{n, \mathrm{R}} ,
      \label{eq:SSE_PS_right}\\
    \hat{H}^\ast \psi_{n, \mathrm{L}}(x)
      &= \mu_n^\ast \psi(x)_{n, \mathrm{L}} .
      \label{eq:SSE_PS_left}
  \end{align}
\end{subequations}%
By comparing \cref{subeqs:SSE_PS} and their complex conjugates we find that right and left eigenfunctions can be expressed by the same function
\begin{equation}
  \psi_{n, \mathrm{R}}(x)
    = \psi^\ast_{n, \mathrm{L}}(x)
    \equiv \psi_n(x) .
  \label{eq:r_eq_l}
\end{equation}
By inserting $\hat{H}^\dagger = \hat{H}^\ast$ into \cref{subeqs:quasi_herm} we find that $\hat{\eta}_\mathrm{R}$ and $\hat{\eta}_\mathrm{L}$ can also be expressed by the same operator
\begin{equation}
  \hat{\eta}_\mathrm{L}
    = \hat{\eta}_\mathrm{R}^\ast
    \equiv \hat{\eta} .
  \label{eq:one_eta}
\end{equation}
In the next section we will apply this theory to investigate under which circumstances a non-Hermitian Hamiltonian with a complex potential can have real or pairs of complex conjugate eigenvalues.

\subsection{Symmetrized multi-well systems}%
\label{subsec:method}

\begin{figure}[tb]
  \includegraphics[width=\columnwidth]{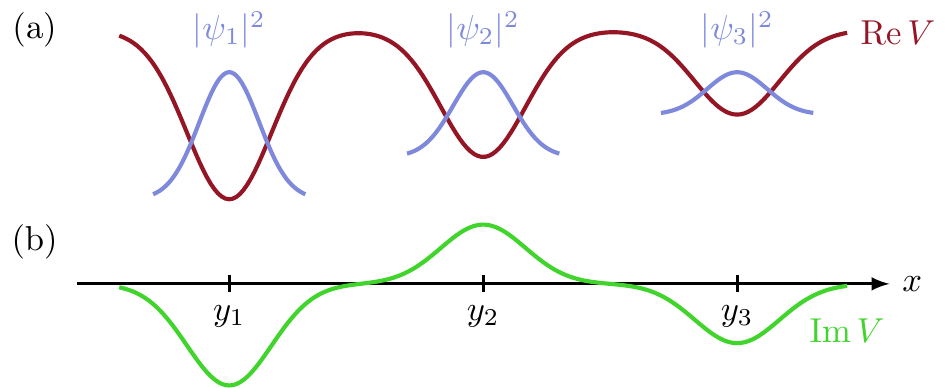}
  \caption{Sketch of a triple-well system. (a) Real part of the potential with the particle densities in each well. (b) The imaginary part of the potential corresponds to the gain-loss profile.}%
  \label{fig:fig1}%
\end{figure}

As mentioned in \cref{subsec:introduction}, we are especially interested in balanced gain and and loss in complex multi-well potentials with either only gain or loss in each well. For this purpose we consider a complex $N$-well potential consisting of Gaussian functions,
\begin{equation}
  V(x) = \sum_{n=1}^{N} \qty(V_n + \imag\Gamma_n)
    \exp\qty(-\frac{(x - a_n)^2}{2\sigma_n^{2}}) .
  \label{eq:multi_well_potential}
\end{equation}
Here, $V_n$, $\Gamma_n$, $\sigma_n$, and $a_n$ are the well depth, the gain-loss parameter, the width, and the position of the center of the $n$-th well, respectively. \Cref{fig:fig1} shows a sketch of the potential \labelcref{eq:multi_well_potential} for the case $N=3$.

Our goal is now to determine the parameters of the potential \labelcref{eq:multi_well_potential} in such a way that at least the first $N$ energy eigenvalues of the corresponding Hamiltonian are real or emerge in complex conjugate pairs. On the one hand, the naive approach would be to find $N$ functions of the first $N$ energy eigenvalues that become zero if the eigenvalues are real or emerge in complex conjugate pairs and afterwards perform a root search for these expressions with respect to the potential parameters. However, this method requires an initial guess for the potential parameters and converges only if this is already close to an actual solution. On the other hand, one could apply the formalism introduced in \cref{subsec:non_herm} to solve the problem. However, this requires the construction of an operator $\hat{\eta}$ satisfying \cref{subeqs:quasi_herm}, which is a very hard problem for extended systems. To circumvent this issue we use the fact that the $N$-well system can be approximately described by a matrix model in which all operators are described by $(N\times N)$-matrices, so that the symmetrization operators can be calculated readily \cite{Dizdarevic2019}. The matrix model is given by
\begin{subequations}
  \begin{equation}
    \vec{H}_\mathrm{eff} \vec{d}_\mathrm{eff} = \mu \vec{d}_\mathrm{eff} ,
    \label{eq:matrix_equation_final_form_0}
  \end{equation}
  where
  \begin{equation}
    \vec{H}_{\mathrm{eff}}
    =  \mqty(\varepsilon_1 + \imag \gamma_1 & -J & & \\
      -J & \ddots & \ddots & \\
      & \ddots & \ddots & -J \\
      & & -J & \varepsilon_N + \imag \gamma_N)
    \label{eq:H_eff}
  \end{equation}
\end{subequations}
with the on-site energies $\varepsilon_n$, the gain-loss terms $\gamma_n$, and the tunneling rates $J$. A detailed derivation of \cref{eq:matrix_equation_final_form_0} can be found in the appendix. If the matrix model is a good approximation of the continuous model, then its energy eigenvalues should be roughly the same as the first $N$ eigenvalues of the continuous system. This is the case if the wells are strongly localized and if they have only a small overlap, \ie{}, if they are deep and narrow. In this case the parameter $J$ does only weakly depend on the parameters $V_n$ and $\Gamma_n$. Furthermore, changing the values of $V_n$ almost only affects $\varepsilon_n$ and changing the values of $\Gamma_n$ almost only affects $\gamma_n$ as long as the changes are small enough. The overlap must however not be too small, as balanced gain and loss is only possible if particles can be exchanged between the wells.

For the $(2\times2)$-matrix model one can explicitly show that one real energy eigenvalue exists if $|\gamma_1||\gamma_2|\le J^2$ holds and if $\gamma_1$ and $\gamma_2$ have opposite signs. Furthermore, on-site energies and gain-loss terms have to be related according to
\begin{equation}
  \varepsilon=\pm(\gamma_1+\gamma_2)\sqrt{-\frac{\gamma_1\gamma_2+J^2}{\gamma_1\gamma_2}},
  \label{eq:2x2_matrix_model}
\end{equation}
where $\varepsilon=\varepsilon_2-\varepsilon_1$. For the $(3\times3)$-matrix model, the eigenvalues are real or emerge in complex conjugate pairs if
\begin{subequations}%
  \label[pluralequation]{subeqs:3x3_matrix_model}
  \begin{align}
    &\varepsilon_1\lessgtr\varepsilon_2\lessgtr\varepsilon_3,
    \label{eq:3x3_matrix_model_eps}\\
    &\gamma_{1,3}\gtrless0,\,\gamma_2\lessgtr0.
    \label{eq:3x3_matrix_model_gam}
  \end{align}
\end{subequations}%

If the spectrum of the matrix model consists of real and pairs of complex conjugate energies we expect that the continuous model possesses as well at least $N$ real or pairwise complex conjugate eigenvalues. For the matrix model the Hilbert space has a finite dimension and one can immediately find the potential parameters for the symmetrized Hamiltonian \cite{Dizdarevic2019}. Our approach will thus be as follows. We chose a configuration of the continuous model, which roughly resembles a configuration in the matrix model with real and pairs of complex conjugate eigenvalues. Then, we manually tune the parameters of the continuous model in such a way that the corresponding matrix model has real or complex conjugate energy eigenvalues. Finally, we perform a root search with respect to the parameters of the continuous potential \labelcref{eq:multi_well_potential}, taking the parameter values obtained from the matrix model as an initial guess, so that the first $N$ energy eigenvalues become either real or pairwise complex conjugate. The root search is performed by the \texttt{minpack hybrid1} routine \cite{minpack} using a modified Powell method \cite{vetterling1989numerical}. The Schr\"odinger equation is solved by the routine \texttt{tridag} \cite{vetterling1989numerical}, which was slightly modified for complex numbers.

To put it in a nutshell: We exploit the symmetrizability of the $N$-dimensional matrix model to get proper initial values to perform a root search of the extended system, so that the first $N$ energy eigenvalues are either real or pairwise complex conjugate.

We want to point out that the symmetrization operators $\hat{\eta}_\mathrm{R}$ and $\hat{\eta}_\mathrm{L}$ can be constructed from the $N$ eigenstates corresponding to the real and complex conjugate energies we find by this method according to \cref{subeqs:etas}. In this case all other states are elements of the kernels of the symmetrization operators, so that $\hat{H}$ is semi-symmetrizable. That is, the continuous model is symmetrizable on the subspace spanned by the first $N$ eigenstates of $\hat{H}$, which correspond to the $N$ eigenvectors of $\vec{H}_{\mathrm{eff}}$, i.\,e.\ $\vec{H}_{\mathrm{eff}}$ satisfies \cref{subeqs:quasi_herm} with the matrix approximations of $\hat{\eta}_{\mathrm{R}}$ and $\hat{\eta}_{\mathrm{L}}$.

\section{Results}%
\label{sec:results}

\subsection{Double-well potential}

\begin{figure*}[tb]
   \includegraphics[width=\textwidth]{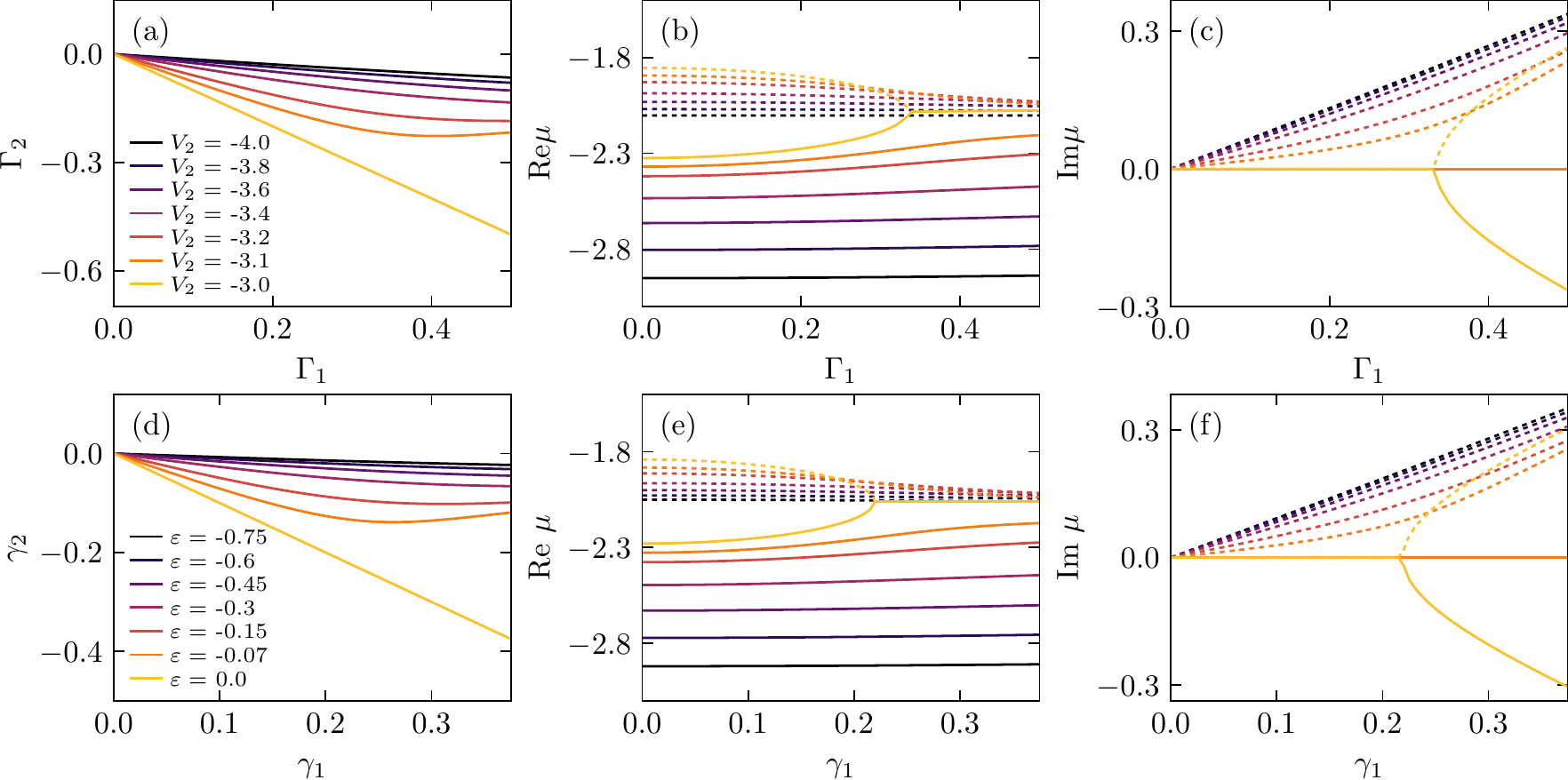}
   \caption{(a) $\Gamma_2$ as a function of $\Gamma_1$ with a real ground state energy for the complex asymmetric double-well potential \labelcref*{eq:multi_well_potential} with $V_1 = -3$, $\sigma_1 = \sigma_2 = 1$, and $a_1 = -a_2 = -1.5$ for different values of $V_2$. (b) Real parts and (c) imaginary parts of the first two energy eigenvalues along these lines. The ground state energies are always real (solid lines), while the energies of the first excited state are always complex (dashed lines) with the exception of the $\mathcal{PT}$-symmetric case $V_1 = V_2 = -3$. (d) $\gamma_2$ as a function of $\gamma_1$ with real ground state energy and (e) the real and (f) the imaginary parts of the spectra in the corresponding matrix model. We find a good agreement between both models. For reference some numerical values are shown in \cref{tab:tab1}.}%
   \label{fig:fig2}%
\end{figure*}

\begin{figure}[tb]
  \includegraphics[width=\columnwidth]{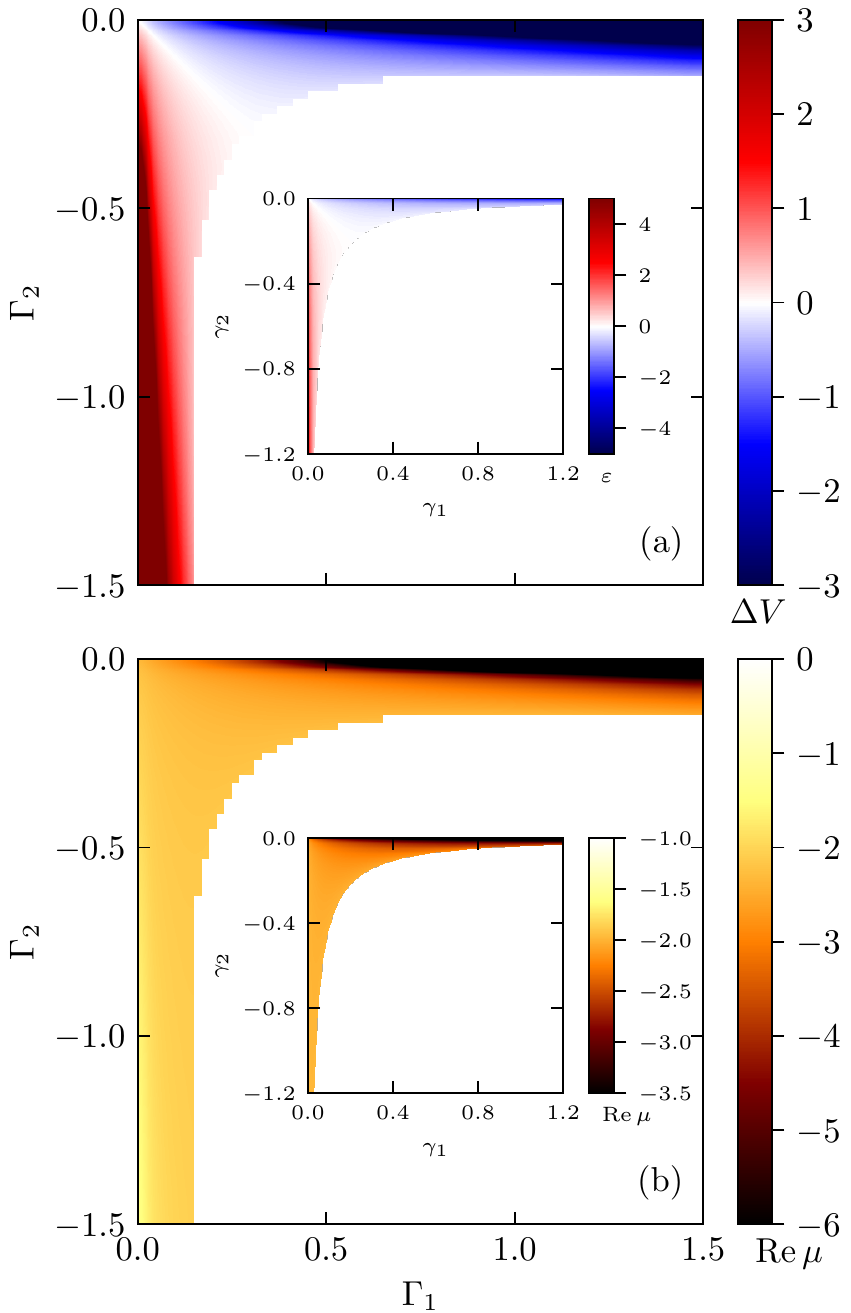}
  \caption{(a) Values of $\Delta V(\Gamma_1, \Gamma_2)$ with a real ground state for the complex asymmetric double-well potential \labelcref*{eq:multi_well_potential} with $V_1 = -3$, $\sigma_1 = \sigma_2 = 1$, and $a_1 = -a_2 = -1.5$ as well as (b) the corresponding real ground state energy $\mu_1(\Gamma_1, \Gamma_2)$. The insets show the corresponding quantities in the matrix model, namely (a) the difference of the onsite energies $\varepsilon(\gamma_1, \gamma_2)$ and (b) the real ground state energy $\mu(\gamma_1, \gamma_2)$. Again, we find good agreement between both models.}%
  \label{fig:fig3}%
\end{figure}

\begin{figure}[tb]
  \includegraphics[width=\columnwidth]{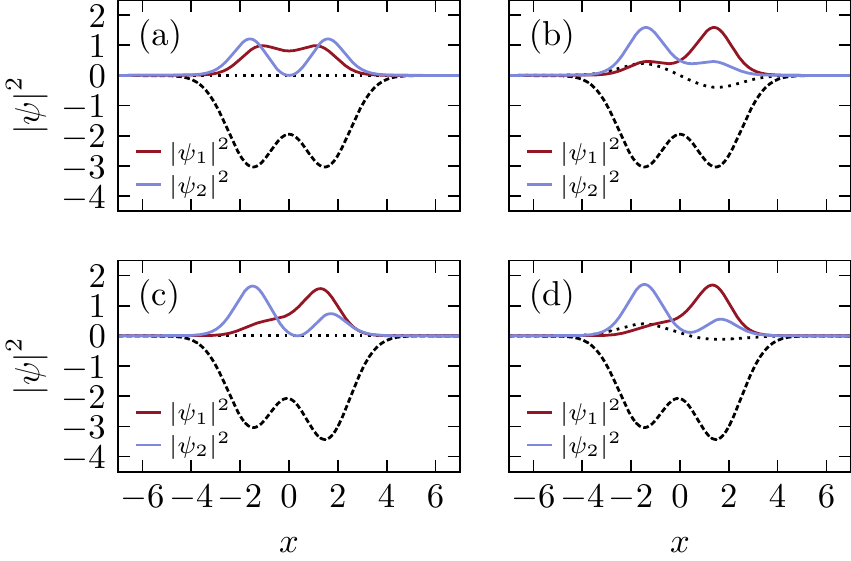}
  \caption{Absolute squares of the wave functions of the complex asymmetric double-well potential \labelcref*{eq:multi_well_potential} (real/imaginary part shown dashed/dotted) with $V_1 = -3$, $\sigma_1 = \sigma_2 = 1$, $a_1 = -a_2 = -1.5$ with a real ground state in the case of (a) a real symmetric, (b) a complex $\mathcal{PT}$-symmetric, (c) a real asymmetric, and (d) a complex asymmetric potential. In the asymmetric cases  $V_2=-3.4$ holds.}%
  \label{fig:fig4}%
\end{figure}

We now want to apply the method discussed in \cref{subsec:method} to a complex double-well potential of the form \labelcref*{eq:multi_well_potential} with $N = 2$. It remains then to investigate if and for which values of the potential parameters such a system can posses real or pairs of complex conjugate energies.

It was shown in Ref.~\cite{Dizdarevic2019} that more than one real energy eigenvalue can only be found if the potential is $\mathcal{PT}$-symmetric, \ie{}, its real part is symmetric and its imaginary part is antisymmetric. To find the range of the potential parameters for which one real energy eigenvalue exists, we set $V_1=-3$, $\sigma_1 = \sigma_2=1$, and $a_2 = -a_1 = 1.5$. We then choose different fixed values for $V_2$ between $-4$ and $-3$. For every value of $V_2$  we vary $\Gamma_1$ between $0$ and $0.5$ and determine the value of $\Gamma_2$ for which the ground state energy becomes real. For this we perform a one-dimensional root search of the imaginary part of the ground state energy with respect to $\Gamma_2$. This root search is simple enough so that no initial guess has to be determined through the matrix model yet.

A comparison with the matrix model further requires that we find the corresponding parameters. For the parameter $J$ we take the value for $\Gamma_1 = \Gamma_2 = 0$ and $V_1 = V_2 = -3$ given by $J = 0.21918847$. For every combination of $V_1$ and $V_2$, the values of $\varepsilon_1$ and $\varepsilon_2$ are calculated for $\Gamma_1=\Gamma_2 = 0$. The minimum and maximum values for $\gamma_1$ are taken from the case with $V_1 = V_2 = -3$, which yields $\gamma_1 \in [0, 0.375]$, while $\gamma_2$ is calculated by the condition that the ground state energy has to be real.

The results are summarized in \cref{fig:fig2,tab:tab1}. \Csubrefrange{fig:fig2}{a}{c} show the results of the continuous model, while \csubrefrange{fig:fig2}{d}{f} show the results of the matrix model. We find an excellent agreement between both models. For all displayed values $V_2 < V_1$ holds. \Csubref{fig:fig2}{d} further shows that $\abs{\Gamma_2} < \abs{\Gamma_1}$ holds and that the value of $\abs{\Gamma_2}$ decreases as the well described by $V_2$ becomes deeper. This is because for a deeper well the amplitude of the wave function increases. To compensate for this a smaller $\Gamma_2$ is required. Regarding the energies, there is always a real ground state energy and a complex excited state energy with an imaginary part growing with increasing gain and loss terms. Remarkably, balanced gain and loss is possible for larger values of $\Gamma_1$ in the asymmetric system than in the $\mathcal{PT}$-symmetric system.

Next, we want to find a whole parameter range with a real ground state energy. For this purpose we vary $\Gamma_1$ and $\Gamma_2$ on a lattice with $0 \le \Gamma_1 \le 1.5$ and $-1.5 \le \Gamma_2 \le 0$ in steps of $0.2$. For each lattice point we determine $V_2$ in such a way that the ground state becomes real. We therefore perform a one-dimensional root search of the imaginary part of the ground state energy with respect to $V_2$. For comparison with the matrix model we now have to map the parameter region from the continuous model onto a parameter region in the matrix model again. We find $0 \le \gamma_1 \le 1.18736616$ and $-1.18736616 \le \gamma_2 \le 0$. Finally, we get a starting value for $\varepsilon_1$ by setting $\Gamma_1=\Gamma_2=0$ and $V_1=V_2=-3$, which yields $\varepsilon_1=-1.95524871$. The results can be seen in \cref{fig:fig3}. Here, \csubrefs{fig:fig3}{a}{c} show $\Delta V=V_2-V_1$ and $\varepsilon = \varepsilon_2 - \varepsilon_1$, while \csubrefs{fig:fig3}{b}{d} show the real parts of the ground state energy of the continuous model and the matrix model, respectively.

\begin{table}[tb]
  \caption{Numerical values of the gain and loss parameters $\Gamma_1$ and $\Gamma_2$ and the corresponding first two eigenvalues $\mu_1$ and $\mu_2$ shown in \cref{fig:fig2}.}%
  \label{tab:tab1}%
  \begin{tabularx}{8.6cm}
    {c@{\extracolsep{\fill}}c@{\extracolsep{\fill}}c@{\extracolsep{\fill}}c}
    \toprule
    $\Gamma_1$ & $\Gamma_2$ & $\mu_1$ & $\mu_2$ \\
    \midrule
    $0.0$ & $-0.000000$
      & $-2.419323$
      & $-1.928826$
    \\
    $0.1$ & $-0.058705$
      & $-2.413120$
      & $-1.934277 + 0.033013 \imag$
    \\
    $0.2$ & $-0.112381$
      & $-2.394828$
      & $-1.950414 + 0.070374 \imag$
    \\
    $0.3$ & $-0.154672$
      & $-2.366256$
      & $-1.975736 + 0.117561 \imag$
    \\
    $0.4$ & $-0.178948$
      & $-2.333120$
      & $-2.005015 + 0.180267 \imag$
    \\
    $0.5$ & $-0.185180$
      & $-2.304364$
      & $-2.029659 + 0.258526 \imag$
    \\
    \bottomrule
  \end{tabularx}
\end{table}

We again find an excellent agreement between both models, though the regions with real ground state energies are slightly deformed. In both cases the regions are limited by the $\gamma$-axis, respectively the $\Gamma$-axis, and by a hyperbolic curve. In both models the depths of both wells are equal  along the line where gain and loss terms are equal. This corresponds to the $\mathcal{PT}$-symmetric case, where in principle all bound states could be real. In the area $\abs{\Gamma_1} < \abs{\Gamma_2}$ the value of $\abs{V_2}$ increases and the increase becomes stronger towards the $\Gamma_2$-axis. In the area $\abs{\Gamma_1} < \abs{\Gamma_2}$ the value of $\abs{V_2}$ decreases and the decrease also becomes stronger towards the $\Gamma_1$-axis. This can again be explained by the compensation of gain and loss being necessary for them to be balanced. We also want to point out that $V_2$ can attain on positive values, as can be seen in \cref{fig:fig3}. However, in this case \cref{eq:multi_well_potential} describes no longer a proper double-well potential.

\begin{figure*}[tb]
  \includegraphics[width=\textwidth]{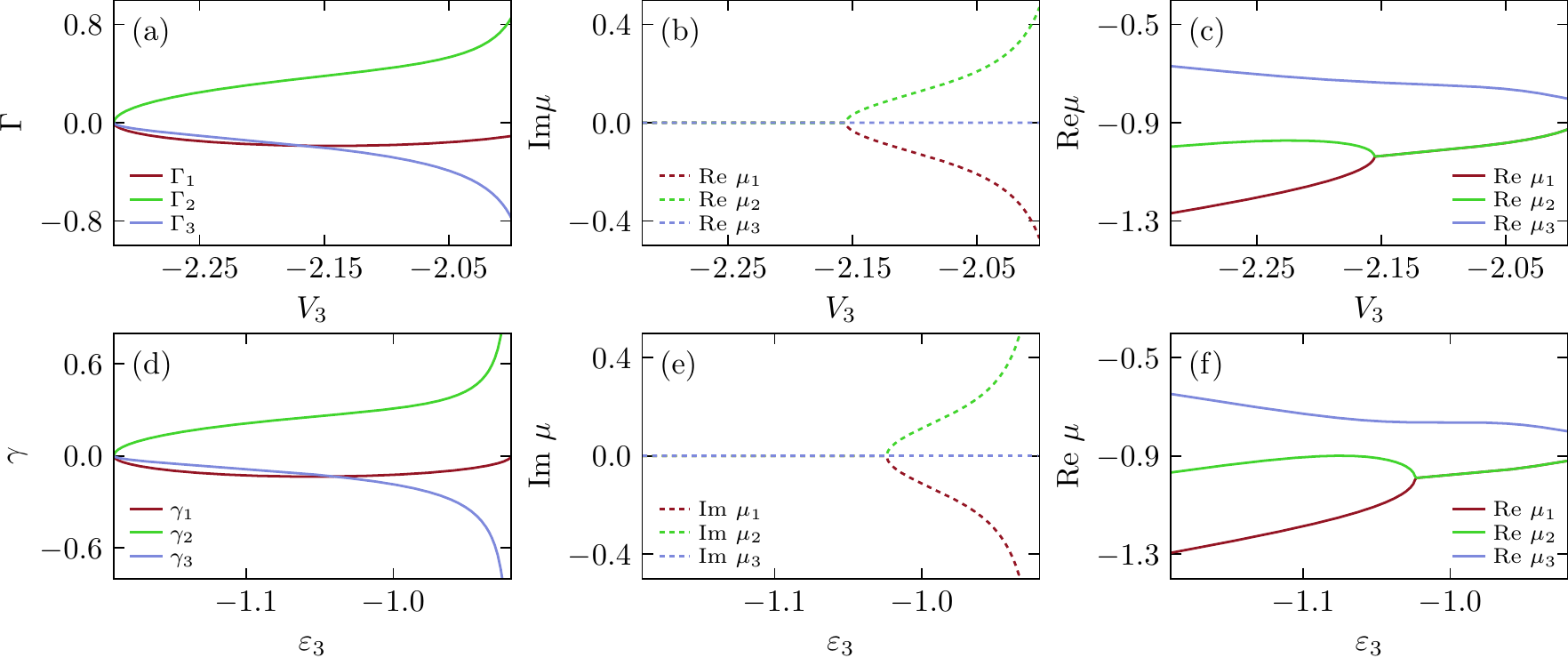}
  \caption{(a) Gain and loss parameters $\Gamma_1$, $\Gamma_2$, and $\Gamma_3$ of the triple-well potential \labelcref*{eq:multi_well_potential} with $\sigma_1 = \sigma_2 = \sigma_3 = 1/\sqrt{2}$, $a_1 = -a_3 = -3$, $a_2 = 0$, $V_1 = -1.8$, and $V_2 = -2$ for balanced gain and loss. (b) Imaginary parts and (c) real parts of the first three energy eigenvalues. (d) Gain an loss parameters $\gamma_1$, $\gamma_2$, and $\gamma_3$, as well as (e) imaginary parts and (f) real parts of the three energy eigenvalues in the corresponding matrix model. We find a good agreement between both models. For reference some numerical values are shown in \cref{tab:tab2}.}%
  \label{fig:fig5}%
\end{figure*}

\begin{table*}[tb]
  \begin{minipage}{\textwidth}
    \caption{Numerical values of the gain and loss parameters $\Gamma_1$, $\Gamma_2$, and $\Gamma_3$ and the corresponding first three eigenvalues $\mu_1$, $\mu_2$, and $\mu_3$ shown in \cref{fig:fig5} for different values of $V_3$.}%
    \label{tab:tab2}%
  \end{minipage}
  \begin{tabularx}{17.8cm}
    {c@{\extracolsep{\fill}}c@{\extracolsep{\fill}}c@{\extracolsep{\fill}}c@{\extracolsep{\fill}}c@{\extracolsep{\fill}}c@{\extracolsep{\fill}}c}
    \toprule
    $V_3$ & $\Gamma_1$ & $\Gamma_1$ & $\Gamma_3$ & $\mu_1$ & $\mu_2$ & $\mu_3$ \\
    \midrule
    $-2.30$
      & $-0.084905$ & $0.132971$ & $-0.051193$
      & $-1.251887$ & $-0.989558$ & $-0.678807$
    \\
    $-2.25$
      & $-0.149449$ & $0.248590$ & $-0.105274$
      & $-1.204718$ & $-0.974868$ & $-0.702382$
    \\
    $-2.20$
      & $-0.178139$ & $0.321408$ & $-0.151567$
      & $-1.147328$ & $-0.974700$ & $-0.722134$
    \\
    $-2.15$
      & $-0.188058$ & $0.381013$ & $-0.203226$
      & $-1.034994 - 0.033549 \imag$
      & $-1.034970 + 0.033544 \imag$
      & $-0.736581 + 0.000004 \imag$
    \\
    $-2.05$
      & $-0.158993$ & $0.532403$ & $-0.389020$
      & $-0.982308 - 0.208872 \imag$
      & $-0.982308 + 0.208872 \imag$
      & $-0.762238 - 0.000000 \imag$
    \\
    $-2.00$
      & $-0.107861$ & $0.851304$ & $-0.769672$
      & $-0.926036 - 0.470621 \imag$
      & $-0.926036 + 0.470624 \imag$
      & $-0.801894 - 0.000003 \imag$
    \\
    \bottomrule
  \end{tabularx}
\end{table*}

Differences between the two models can only be seen in the vicinity of the axes. In the matrix model $\varepsilon$ diverges close to the $\gamma$-axes, while $\abs{V_2 - V_1}$ also increases towards the $\Gamma$-axes for the continuous model. Furthermore, in the matrix model $\varepsilon$ is exactly symmetric with respect to the line $\gamma_2 = -\gamma_1$, which is not the case in the continuous model in the vicinity of the $\Gamma$-axes. To put it in a nutshell, in the vicinity of the gain and loss axes we find differences between both models. One reason for this might be that in the matrix model the parameters $\varepsilon_n$ can no longer be interpreted as well depths and $\varepsilon_n > 0$ still describe bound states by construction. In the continuous model on the other hand $V_n > 0$ is connected with the occurrence of scattering states. However, apart from this there is good agreement between the two models both in terms of the well depths and the ground state energies.

Finally, we want to take a look at the wave functions at specific points in the parameter space. \Cref{fig:fig4} shows the wave functions of the first two states for $V_1 = -3$, $\sigma_1 = \sigma_2=1$, and $a_2 = -a_1 = 1.5$ for different values of $V_2$ and $\Gamma_1$, while $\Gamma_2$ is chosen again in such a way that the ground state energy is real. \Csubrefs{fig:fig4}{a} and (c) show a symmetric real potential with $V_2 = -3$ and an asymmetric real potential with $V_2 = -3.4$, respectively. In the asymmetric potential the ground state wave function has a larger amplitude in the deeper well. The imaginary part of the potential is turned on in \csubrefs{fig:fig4}{b} and (d) with $\Gamma_1 = 0.4$. These values are in the broken $\mathcal{PT}$-symmetric regime, as one can see in \cref{fig:fig2}. Thus, in the $\mathcal{PT}$-symmetric case shown in \csubref{fig:fig4}{b} there exist no real energies. The reason for this is that the loss in the right well cannot compensate for the gain in the left well, which effectively leads to an overall particle increase. \Csubref{fig:fig4}{d} shows the asymmetric complex case, where gain and loss are balanced for the ground state, while they are unbalanced for the excited state.

\subsection{Multi-well potentials}

In the $(N\times N)$-matrix model for $N\ge 3$ one could always find a symmetrization matrix with empty kernel and thus find parameters for which $N$ energy eigenvalues are real or pairwise complex conjugate \cite{Dizdarevic2019}. We therefore expect, that a continuous $N$-well system will as well posses $N$ such energy eigenvalues. We want to examine this explicitly for a triple-well potential of the form \cref{eq:multi_well_potential} with $N=3$. Again we choose fixed values for the well widths and distances, that is, $\sigma_1 = \sigma_2 = \sigma_3 = \flatfrac{1}{\sqrt{2}}$, $a_3 = -a_1 = 3$, and $a_2 = 0$. We then want to take fixed values of $V_1$, $V_2$, and $V_3$ and determine the corresponding gain-loss parameters $\Gamma_1$, $\Gamma_2$, and $\Gamma_3$ for which the first three energy eigenvalues are real or pairwise complex conjugate. In order to do so we have to find three functions of the first three energy eigenvalues, which become zero if the energy eigenvalues are real or emerge in complex conjugate pairs. For example one can easily show, that the spectrum has the required structure if the equations
\begin{subequations}%
  \label{eq:zero_search_3d}%
  \begin{align}
  \Im(\mu_1 + \mu_2 + \mu_3) &= 0 ,
    \label{zero_search_3d_a} \\
  \Im(\mu_1\mu_2 + \mu_1\mu_3 + \mu_2\mu_3) &= 0 ,
    \label{zero_search_3d_b} \\
  \Im(\mu_1 \mu_2 \mu_3) &= 0
    \label{zero_search_3d_c}
  \end{align}
\end{subequations}
are fulfilled. Thus we have to solve \cref{eq:zero_search_3d} with respect to $\Gamma_1$, $\Gamma_2$, and $\Gamma_3$ for fixed values of $V_1$, $V_2$, and $V_3$. Furthermore, we now also need to determine an initial guess of the system parameters by means of the matrix model as described in \cref{subsec:method}. For this we chose $V_1 = -1.8$, $V_2=-2.0$, and $V_3 = -2.2$ and find
\begin{subequations}%
  \label{eq:start_value_triple_gauss_well}%
  \begin{align}
    \Gamma_1 &= -0.178139 , \\
    \Gamma_2 &= 0.321408 , \\
    \Gamma_3 &= -0.151567 .
  \end{align}
\end{subequations}
To find whole parameter ranges with real and pairs of complex conjugate energies, we repeat this process along a grid in the $V_1$-$V_3$-plane, where we keep $V_2=-2.$ From the matrix model we already know that solutions can only exist if either $V_1>V_2$ and $V_3<V_2$ or $V_1<V_2$ and $V_3>V_2$, see \cref{subeqs:3x3_matrix_model}. At every step we change only one of the well depths slightly, so that we can take the solution of the previous step as initial guess for the current step.

\begin{figure}[tb]
  \includegraphics[width=\columnwidth]{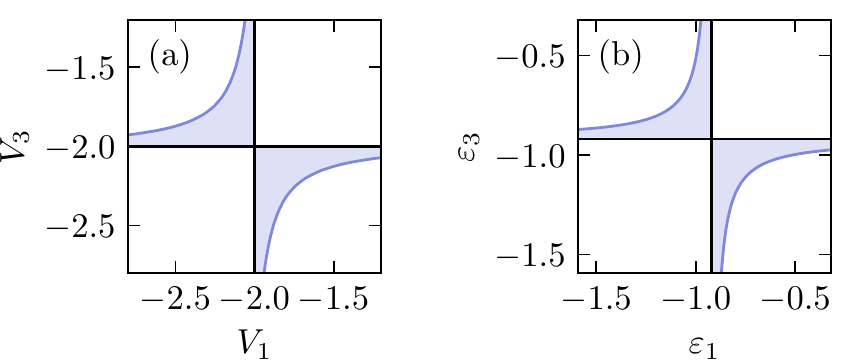}
  \caption{(a) Section of $V_1$-$V_3$-plane of the parameter space of the triple-well potential \labelcref*{eq:multi_well_potential} with $\sigma_1 = \sigma_2 = \sigma_3 = 1/\sqrt{2}$, $a_1 = -a_3 = -3$, $a_2 = 0$, and $V_2=-2$ in which three bound states with real energy eigenvalues exist and (b) the corresponding parameter range in the matrix model. Both models have a good qualitative agreement.}%
  \label{fig:fig6}%
\end{figure}

The results for the triple-well potential are summarized in \cref{fig:fig5,tab:tab2}. \Csubref{fig:fig5}{a} shows the solutions for $\Gamma_1$, $\Gamma_2$, and $\Gamma_3$ in case of $V_1 = -1.8$ for different values of $V_3$. The gain-loss parameters diverge towards $V_3 = -2$ and become imaginary at $V_3 = -2.319$. \Csubrefs{fig:fig5}{b} and (c) show the imaginary and real parts of the first three energy eigenvalues. For $V_3 < -2.15$ all energies are real. At $V_3 = -2.15$ a bifurcation occurs, so that the ground and the excited states form a complex conjugate pair. \Csubrefs{fig:fig5}{d}, (e), and (f) show the corresponding quantities in the matrix model. As for the double-well potential we find an excellent agreement between both models. Starting from the point \labelcref{eq:start_value_triple_gauss_well} we can iteratively calculate solutions in different directions. By connecting the points where the gain-loss parameters become imaginary we can determine the boundary of the section in $V_1$-$V_3$-plane in which real or pairs of complex conjugated energies exist. \Cref{fig:fig6} shows that there is again a good agreement between the matrix model and the continuous model.

For systems with more than three wells we already know from investigations of the matrix model that we can determine the potential parameters in such a way that the first $N$ eigenvalues are real or pairwise complex conjugate, where $N$ is the number of potential wells. Due to the excellent agreement with matrix models, this should in principle also be possible for systems with spatially extended multi-well potentials.

\section{Conclusions}%
\label{sec:conclusions}

In this paper we investigated balanced gain and loss in non-interacting Bose-Einstein condensates in complex asymmetric multi-well potentials described by a Schr\"odinger equation in position space. Gain and loss are effectively described by adding an imaginary part to the potential. If gain and loss are balanced, then the corresponding Hamiltonian has real eigenvalues.

To find the circumstances under which the described system has real eigenvalues, we used the fact that, in case of strongly localized potential wells, the $N$-well system can be well described by an $(N \times N)$-matrix model. By means of this matrix approximation we developed a reliable method to construct complex $N$-well potentials with either only gain or loss in each well, which yields $N$ real or pairs of complex conjugate eigenvalues for $N>2$ and one real eigenvalue for $N=2$. We did this explicitly for double and triple-well systems and found excellent agreement between the continuous model and the matrix model. A peculiarity of the double-well system is that, apart from the $\mathcal{PT}$-symmetric case, only one real eigenvalue can exist. In the matrix model this can be explained by the fact that only one eigenvector is not in the kernel of the symmetrization operator. We thus expect that the Hamiltonian for the continuous double-well potential is also semi-symmetrizable with an operator $\hat{\eta}$ which contains all eigenstate except for one in its kernel.

The presented method should in principle work for every $N$-well potential with only either gain or loss in each well. However, it cannot be used to construct potentials with more than $N$ real or complex conjugate eigenvalues, which would require for a new method. Another limitation is that so far only one-dimensional systems were considered. As any setup for a possible experimental realization is three dimensional, a generalization of the continuous model to three spatial dimensions might be required, though we do not expect any new effects to appear \cite{Dast2012}.

To take the contact interaction between the atoms into account, an analysis of the nonlinear system is required, which could allow for interesting applications (\eg{}, see Ref.~\cite{Assawaworrarit2017}). Last but not least the concept used here could also be applied to many-body systems beyond the mean-field limit, where gain and loss are necessarily asymmetric \cite{Dast2014}.

% ---------------------------------- Appendix ---------------------------------

% Specify following sections are appendices. Use \appendix* if there
% only one appendix.
%\appendix
%\section{}

\appendix*

\section{Derivation of the matrix model}%
\label{app:deriv_matrix_model}

In this appendix we show the derivation of the matrix model \labelcref{eq:matrix_equation_final_form_0} starting from the continuous model. For this we discretize the Schr\"odinger equation by approximating the wave function first using a linear combination of the ground states of the single wells without imaginary parts and by integrating them afterwards.

We start the derivation of the matrix model with the Schr\"odinger equation
\begin{equation}
    \hat{H}(x) \psi_\mathrm{ex}^{(l)}(x)
      = E_\mathrm{ex}^{(l)} \psi_\mathrm{ex}^{(l)}(x)
    \label{eq:exact_SE}
\end{equation}
with the exact $l$-th eigenfunction $\psi_\mathrm{ex}^{(l)}(x)$ with eigenvalue $E_\mathrm{ex}^{(l)}$. To derive the matrix model from \cref{eq:exact_SE} we approximate the $l$-th eigenfunction by
\begin{equation}
    \psi_\mathrm{an}^{(l)}(x) = \sum_{n=1}^{N} c_n^{(l)} \phi_n(x)
    \label{eq:ansatz_for_matrix_model}
\end{equation}
with the coefficients $c_n^{(l)} \in \mathbb{C}$. The functions $\phi_n(x)$ can in principle be chosen arbitrarily as long as \cref{eq:ansatz_for_matrix_model} is a good approximation for the exact eigenfunction in the sense that $\psi_\mathrm{ex}^{(l)}(x) = \psi_\mathrm{ansatz}^{(l)}(x) + \delta\psi^{(l)}(x)$ with
\begin{equation}
    \abs{\frac{\delta\psi^{(l)}(x)}{\psi_\mathrm{ex}^{(l)}(x)}}
    \ll 1\ \forall \, x \in \mathbb{R} .
    \label{eq:good_approx}
\end{equation}
Inserting \cref{eq:ansatz_for_matrix_model} into \cref{eq:exact_SE} yields
\begin{equation}
    \hat{H}(x) \psi_\mathrm{an}^{(l)}(x)
      = E_\mathrm{ex}^{(l)} \psi_\mathrm{an}^{(l)}(x)
        + \qty(E_\mathrm{ex}^{(l)} - \hat{H}(x))
      \delta\psi^{(l)}(x) .
    \label{eq:inserting_into_SE}
\end{equation}
Now we multiply both sides of \cref{eq:inserting_into_SE} with $\phi_m^\ast(x)$ and integrate over $\mathbb{R}$ to obtain the matrix equation
\begin{equation}
    \vec{H} \vec{c}^{(l)}
      = E_\mathrm{ex}^{(l)} \vec{K} \vec{c}^{(l)}
      + \vec{\xi}^{(l)}
    \label{eq:matrix_equation_perturbed}
\end{equation}
with the matrix elements
\begin{subequations}%
  \label{subeqs:matrix_elements}%
  \begin{align}
    H_{mn} &= \int_{\mathbb{R}} \dd{x}
        \dv{\phi_m^\ast(x)}{x} \dv{\phi_n(x)}{x} \nonumber \\
      &\quad+ \int_{\mathbb{R}} \dd{x}
        \phi_m^\ast(x) V_\mathrm{ext}(x) \phi_n(x) ,
      \label{eq:H_mn} \\
    K_{mn} &= \int_{\mathbb{R}} \dd{x}
      \phi_m^\ast(x) \phi_n(x) ,
      \label{eq:K_mn} \\
    \xi_{m}^{(l)} &= \int_{\mathbb{R}} \dd{x}
        \phi_m^\ast(x) \qty(E_\mathrm{ex}^{(l)} - \hat{H}(x))
          \delta\psi^{(l)}(x)
        \nonumber \\
      &= \int_{\mathbb{R}} \dd{x}
        \phi_m^\ast(x) \qty(E_\mathrm{ex}^{(l)} - V(x)) \delta\psi^{(l)}(x)\nonumber\\
        &\quad + \int_{\mathbb{R}} \dd{x}
          \dv{\phi_m^\ast(x)}{x} \dv{\delta\psi^{(l)}(x)}{x} .
        \label{eq:chi_m}
  \end{align}
\end{subequations}%
It is important to note that $E_\mathrm{ex}^{(l)}$ in \cref{eq:matrix_equation_perturbed} is still the exact energy of the continuous system and that \cref{eq:matrix_equation_perturbed} itself is exact.
Because of \cref{eq:good_approx} it is reasonable to assume that $\vec{\xi}^{(l)}$ is small and thus negligible, so that there exist $\mu^{(l)} \approx E_\mathrm{ex}^{(l)}$ and $\vec{d}^{(l)} \approx \vec{c}^{(l)}$ satisfying the generalized eigenvalue problem
\begin{equation}
    \vec{H} \vec{d}^{(l)} = \mu^{(l)} \vec{K} \vec{d}^{(l)} .
    \label{eq:matrix_equation_unperturbed}
\end{equation}
\Cref{eq:matrix_equation_unperturbed} can be transformed to an ordinary eigenvalue problem with the method of symmetric orthogonalization. For this we introduce the matrix
\begin{equation}
    \vec{X} = \vec{U}^\dagger \vec{D}^{-\frac{1}{2}} \vec{U}
    \label{eq:matrix_X}
\end{equation}
with $\vec{D}$ being the diagonal matrix of eigenvalues of $\vec{K}$ and $\vec{U}$ being the unitary matrix of the corresponding eigenvectors. With
\begin{subequations}%
    \label{subeqs:eff_matrices}
    \begin{align}
        \vec{H}_\mathrm{eff} &= \vec{X} \vec{H} \vec{X} ,
          \label{eq:H_eff_app}\\
        \vec{c}_\mathrm{eff} &= \vec{X}^{-1} \vec{c}
          \label{eq:c_eff_app}
    \end{align}
\end{subequations}%
we finally arrive at the matrix model represented by the ordinary eigenvalue equation
\begin{equation}
    \vec{H}_\mathrm{eff} \vec{d}_\mathrm{eff} = \mu \vec{d}_\mathrm{eff} .
      \label{eq:matrix_equation_final_form}
\end{equation}

% If you have acknowledgments, this puts in the proper section head.
%\begin{acknowledgments}
% put your acknowledgments here.
%\end{acknowledgments}

% Create the reference section using BibTeX:
% ****** Start of file paper.bbl ****** %
%merlin.mbs apsrev4-1.bst 2010-07-25 4.21a (PWD, AO, DPC) hacked
%Control: key (0)
%Control: author (72) initials jnrlst
%Control: editor formatted (1) identically to author
%Control: production of article title (-1) disabled
%Control: page (0) single
%Control: year (1) truncated
%Control: production of eprint (0) enabled
%
% ****** End of file paper.bbl ****** %

\end{document}